\documentclass[11pt]{article}
\usepackage[utf8]{inputenc}
\usepackage{amssymb,amsmath,amsthm}
\usepackage[english]{babel}
\usepackage[T1]{fontenc}
\usepackage[dvipsnames]{xcolor}
\usepackage{appendix} 
\usepackage{bm}
\usepackage{lipsum}
\usepackage{hyperref}
\usepackage{graphicx}
\usepackage{siunitx}
\hypersetup{colorlinks,linkcolor={blue},citecolor={blue},urlcolor={blue}} 
\usepackage[authoryear]{natbib}
\usepackage{gensymb}
\usepackage{booktabs}
\usepackage{multirow}
\usepackage{siunitx}
\usepackage{bbm}
\usepackage[absolute,overlay]{textpos}
\def\pr{\text{Pr}}%

\def\R{\mathbb{R}}

\def\xx{\mathbf{x}}
\def\yy{\mathbf{y}}
\def\zz{\mathbf{z}}
\def\ww{\mathbf{w}}
\def\pr{\text{Pr}}%

\numberwithin{equation}{section}
\theoremstyle{plain}

\theoremstyle{remark}

\theoremstyle{remark}

\def\R{\mathbb{R}}

\usepackage[total={170mm,230mm},
 left=25mm,
 top=20mm]{geometry}
 
\newgeometry{top=5mm,left=25mm,right=25mm,bottom=25mm} 

\title{Practical strategies for GEV-based regression models for extremes}

\author{Daniela Castro-Camilo$^{1*}$, Rapha\"el Huser$^2$, and H{\r{a}}vard Rue$^2$}

\footnotetext[1]{
\baselineskip=10pt School of Mathematics and Statistics, University of Glasgow, UK.}
\footnotetext[2]{ 
\baselineskip=10pt Computer, Electrical and Mathematical Sciences and Engineering (CEMSE) Division, King Abdullah University of Science and Technology (KAUST), Thuwal, Saudi Arabia.}

\date{May 6, 2022}

\begin{document}

\maketitle

\begin{center}
{\large{\bf Abstract}}
\end{center}
The generalised extreme value (GEV) distribution is the only possible limiting distribution of properly normalised maxima of a sequence of independent and identically distributed random variables.
As such, it has been widely applied to approximate the distribution of maxima over blocks.
In these applications, GEV properties such as finite lower endpoint when the shape parameter $\xi$ is positive or the loss of moments due to the magnitude of $\xi$ are inherited by the finite-sample maxima distribution.
The extent to which these properties are realistic for the data at hand has been widely ignored.
Motivated by these overlooked consequences in a regression setting, we here make three contributions.
First, we propose a blended GEV (bGEV) distribution, which smoothly combines the left tail of a Gumbel distribution (GEV with $\xi=0$) with the right tail of a Fr\'echet distribution (GEV with $\xi>0$). Our resulting distribution has, therefore, unbounded support. 
Second, we proposed a principled method called property-preserving penalised complexity (P$^3$C) priors to decide on the existence of the GEV distribution first and second moments a priori.
Third, we propose a reparametrisation of the GEV distribution that provides a more natural interpretation of the (possibly covariate-dependent) model parameters, which in turn helps define meaningful priors.
We implement the bGEV distribution with the new parameterisation and the P$^3$C prior approach in the R-INLA package to make it readily available to users.
We illustrate our methods with a simulation study that reveals that the GEV and bGEV distributions are comparable when estimating the right tail under large-sample settings. Moreover, some small-sample settings show that the bGEV fit slightly outperforms the GEV fit. Finally, we conclude with an application to NO$_2$ pollution levels in California that illustrates the suitability of the new parametrisation and the P$^3$C prior approach in the Bayesian framework.

\par\vfill\noindent

\restoregeometry 
\newgeometry{top=20mm,left=25mm,right=20mm,bottom=20mm}
\baselineskip=20pt

\newpage
\section{Introduction}~\label{sec:introduction}
The generalised extreme value distribution (GEVD) is a three {parameter} family that describes the asymptotic behaviour of properly renormalised maxima of a sequence of independent and identically distributed random variables. 
It is parametrised in terms of a location $\mu\in\R$, a scale $\sigma>0$ and a shape parameter $\xi\in\R$ that controls the type of tail of the GEVD.
Indeed, if $\xi$ is zero, the GEVD has unbounded (parameter-free) support, whereas if $\xi$ is positive, the limiting distribution is heavy-tailed with infinite upper endpoint but finite lower endpoint that depends on the parameters.
If $\xi<0$, then the {short}-tailed GEVD has a finite upper endpoint that also depends on the parameter.
We can see then that when $\xi\neq0$, the GEVD
does not obey the general regularity conditions for likelihood-based inference~\citep{stuart2004kendall}{; see \citet{Smith:1985} for an in-depth study of maximum likelihood inference for a class of irregular models that includes the GEVD.} 
Moreover, by using the {asymptotic} GEVD as an approximation for the distribution of maxima over {finite} blocks, GEV properties, such as finite lower bound in the case $\xi>0$ and finite upper bound in the case $\xi<0$, are inherited by the original maxima distribution, which might not have bounded support. 

{Despite the above features, many} authors have shown the practical usefulness of the GEV family as an approximation to the distribution of block maxima. 
For instance, \cite{broussard1998behavior} use the GEV {distribution} to analyse the behaviour of extreme losses in a German stock index, whereas \cite{bruun1998comparison} assume a GEVD to estimate the probability of coastal flooding. 
Extensions using regression models to include covariate effects have been used in~\cite{el2007generalized}, \cite{el2009joint}, and \cite{cannon2010flexible}, among others.
Flexible parameter smoothing using the class of vector generalised linear and additive models was proposed in~\cite{yee2007vector}{, and this framework was recently exploited by \citet{zhong2022modeling} for an application to heatwave hazard assessment in Europe}. 
Further extensions to combine spatial information across locations have been analysed in \cite{casson1999spatial}, while~\cite{westra2013global} studied the effect of climate change on global precipitation annual maxima{, and \citet{Johannesson.etal:2021} embedded space-time covariates and spatially correlated random effects within GEV parameters to predict extreme river flows over the whole U.K.\ territory.}
 Additionally, \cite{huang2016estimating} conducted a detailed study on the usefulness and potential limitations of the
GEVD to model seasonal maximum and minimum temperatures from a millennial-scale climate simulation.
From a theoretical perspective, \cite{stein2017should} introduces two new results on the limiting distribution of block maxima, motivated by the study of annual temperature extremes.
The need for such new results comes from the upper bound imposed by the GEV when it is used as an approximation to the distribution of properly renormalised annual maximum temperatures.
Indeed, \cite{stein2017should} argues that
the existence of an upper bound means that for the GEVD to be a suitable approximation of the annual maximum temperature distribution, we would need a theoretical result that applies to the maximum of a large number of random variables with varying upper bounds.
To solve this artefact created by the GEVD, \cite{stein2017should} develops a framework based on triangular arrays,
where the GEVD appears as a particular case.

As in \cite{stein2017should}, here we focus on an {artefact} created by the GEV {distribution} when it is used as an approximation to the distribution of {finite block} maxima.
Our focus, though, is on the case where the shape parameter is  positive and therefore, the limiting GEV distribution has a lower bound.
To illustrate this artefact, consider $X_1,\ldots, X_n$ independent and identically distributed (i.i.d.) according to a standard Cauchy distribution $F$, which has support in $\R$. 
Then, if $M_n = \max\{X_1,\ldots,X_n\}$ and $M_n^\star = (M_n-b_n)/a_n$ with sequences $a_n>0$ and $b_n\in\R$, one has that
$\pr(M_n\leq z) = F^n(z)$ and $\pr(M_n^\star\leq z) = F^n(a_nz+b_n)$ {(where $a_n$ and $b_n$ can be chosen as $a_n=n\pi^{-1}$ and $b_n=0$)}.
This implies that both $M_n$ and its standardised version $M_n^\star$ have also support in $\R$. 
It can be shown (see, e.g., \citealp{schmiedt2016domains}) that in this case, $\pr(M_n^\star\leq z)\to G(z)=\exp(-1/z)$, $z>0$, as $n\to\infty$, i.e., properly {renormalised} maxima of a sequence of i.i.d.~standard Cauchy variables with support in $\R$ converge to {a unit} {Fr\'echet} random variable, which has support in $(0,\infty).$ 
Therefore, the GEV limit imposes an artificial bounded support to the finite-sample maxima of Cauchy random variables, whose actual finite-$n$ distribution may have non-negligible mass over $(-\infty,0)$. 
{Note that many other similar theoretical examples can be found to illustrate the same mismatch between the finite-$n$ and asymptotic supports of block maxima (e.g., maxima of exponential, beta, or Weibull random variables, just to name a few examples).} 

The artificial bound imposed by the GEV approximation is particularly troublesome when the GEV parameters are covariate-dependent, i.e., $\yy(\xx)\sim\text{GEV}(\mu(\xx), \sigma(\xx), \xi(\xx))$, for $\xx\in\R^d$, $d\geq 1$. 
In practical applications, the elements of $\yy(\xx)$ are usually constructed as the block-maxima of a certain variable, say $\zz(\xx)$, for a given block size.
In this case, the GEV support, $\{z: 1+\xi(\xx)\{z-\mu(\xx)\}/\sigma(\xx)>0\}$, is also covariate-dependent. 
In particular, when $\xi(\xx)>0$, then the lower bound is $\mu(\xx)-\sigma(\xx)/\xi(\xx)$.
A GEVD fitted to data $\yy$ implies that the maximum possible value of $\zz$ should also have a lower bound for any configuration of $\xx$, even those that have not been observed. Moreover, this lower bound is changing with the covariates that have been observed.
Currently, there is no theoretical guarantee that a new configuration of covariates outside the sample support would not predict a value $y^\star$ below the fitted lower bound.
In practice, if we wish to predict with covariates outside the sample support, then the whole model is refitted to account for $y^\star$.
Furthermore, note that if $y^\star$ is observed, this single observation will control the estimates and consequently the lower bound, reducing the importance of the remaining observations.
Therefore, following the reasoning of \cite{stein2017should}, we could argue that for the GEV to be a reasonable model for $\yy$, we would need, in principle, a result that applied to maxima of a large number of random variables with varying lower bound.

The examples presented here illustrate that when the GEVD with positive shape is scaled back to be used as an approximation for finite-sample maxima, it introduces a purely theoretical, unrealistic and inconvenient artefact. Furthermore, while this artefact may be considered negligible for the actual model being fitted, it can have major implications for prediction and computing in regression settings.
We address this issue by modifying the lower tail of the GEVD with positive shape.
Specifically, we construct a blended GEV (bGEV) distribution with support in the whole real line,
based on a combination of the Gumbel ($\xi=0$) and the Fr\'echet ($\xi>0$) distributions.
Our approach shares similarities with spliced models (see, e.g,.~\citealt{reynkens2017modelling,castro2019spliced, opitz2018inla}) and mixture models for extremes~\citep{behrens2004bayesian,do2011regression, carreau2009hybrid,de2004data,frigessi2002dynamic,leonelli2020semiparametric} in the sense that all of these methods join two distributions.
But besides this, our approach is fundamentally different from that of spliced or mixture models.
First of all, mixture models blend two or more distributions at the density level.
This means that when the mixture weights are not constant in the argument, the definition of the associated distribution function requires the computation of a normalising constant, which may not have a closed-form expression.
Our bGEV model blends the Frech\'et and Gumbel distributions at the distribution level so that the resulting density function can straightforwardly be computed and is automatically normalised.
Second, the mixture and spliced models cited above are proposed for threshold exceedances.
For instance, \cite{castro2019spliced} and \cite{opitz2018inla} propose a blend of a Gamma distribution for the bulk and a generalised Pareto distribution (GPD) for the tail, and their approach incorporate the modelling of the exceedance probability.
\cite{frigessi2002dynamic} provides a continuous transition between the bulk (described using a Weibull distribution) and GPD-tail models, avoiding the threshold selection associated with fitting a generalised Pareto distribution. 
As noted by~\cite{scarrott2012review}, their model is appropriate when there is a lower bound on the support. 
On the contrary, the goal of our bGEV model is to modify the left tail of the GEV distribution with positive shape parameter while keeping its upper tail behaviour.

Note that since the GEVD is the only possible non-degenerate limit to properly {renormalised} maxima, we do not propose a different limiting model. 
Instead, we create a distribution that behaves very closely to the GEVD, bypassing the lower bound problem.
This means that we expect to get similar results for both distributions in terms of parameter estimates and return level estimation, which is confirmed by our first simulation study in Section~\ref{sec:simulation}.
The benefit of our proposed bGEV model is that it is numerically more stable and reliable when we want to make predictions. It is also more statistically meaningful in the sense that the support does not depend on parameters.

The bGEVD is expressed in terms of a new parametrisation of the GEVD, that provides a more meaningful interpretation of the model's parameters.
Indeed, the usual location-scale GEV parametrisation looses interpretability when the mean and standard deviation {do not exist}, which is the case when $\xi\geq 1$ and $\xi\geq 0.5$, respectively. 
Our {proposed} bGEV parametrisation is based on a GEV quantile $q_\alpha$ (location) and a quantile range (spread) $s_\beta = q_{1-\beta/2}-q_{\beta/2}$ ($0<\alpha,\beta<1$) whose existence, unlike the mean and standard deviation, are not influenced by the shape parameter.
A consequence of this is that we have bGEV location and spread parameters over which we can meaningfully assign priors in the Bayesian setting.
In particular, we have that independent priors over the bGEV location and spread parameters induce a {joint} prior distribution for the original location and scale parameters.

The selection of a prior distribution for the bGEV shape parameter motivates an additional contribution of this paper. As mentioned above, the shape parameter has a major role in determining the type of tail of the GEVD. 
It is also key in determining important features of the distribution, such as the existence of moments. 
Indeed, the $k$-th moment exists if and only if $\xi<1/k$~\citep{muraleedharan2011characteristic}. 
In a regression context where $\xi$ might depend on covariates, we might have that, e.g., the first and second moments are not \emph{continuous} as a function of the covariates, i.e., mean and variance exist only for some configuration of the covariates. The consequences of this issue have not received much attention in the literature.
By focusing on these low-order moments, we proposed a principle method to decide a prior on the existence of the first and second moments. Our method is called property-preserving penalised complexity (P$^3$C) prior approach, and {exploits} the framework introduced by \cite{simpson2017penalising} to ensure desired features of the bGEVD.

To summarise, the goal of this paper is threefold: 1) to offer an alternative to the classical GEVD that corrects the left tail, providing a parameter-free support; 2) to propose a reparametrisation of the GEVD in terms of a quantile and a quantile range that eases the interpretation of the model's parameters; and 3) to introduce a principled method to decide the existence of low-order moments a prior.

Our proposed bGEV model is implemented using a hierarchical Bayesian framework in the context of latent Gaussian models, using the new parametrisation and the P$^3$C prior approach.
Marginal posterior {distributions} of interest are computed using the integrated nested Laplace approximation (INLA; \citealp{rue2009approximate}).
The bGEV model is freely available and efficiently implemented in the R-INLA package~\citep{bivand2015spatial,art527}.

The remainder of the paper is organised as follows.
In Section~\ref{sec:reparam} we describe the new reparametrisation of the GEVD, while in Section~\ref{sec:bGEV} we describe our modelling approach {based on the bGEVD}. 
The concept of property-preserving penalised complexity {priors} is introduced in Section~\ref{sec:pppc}.
In Section~\ref{sec:simulation} we present three simulation studies that compare the performance of the GEVD and the bGEVD at estimating return levels and provide insights on the influence of the bGEV model parameters.
Section~\ref{sec:application} presents the improvements implied by our model using NO$_2$ concentrations in California. 
Conclusions and a discussion of our methods are given in Section~\ref{sec:discussion}.

\section{A reparametrisation of the GEV distribution}~\label{sec:reparam}
The usual GEV parametrisation in terms of location, scale and shape parameters is {somewhat} convenient as it resembles well-known location-scale families. 
Within these families, we usually associate the location and scale parameters with the mean and standard deviation. 
Nonetheless, in skewed distributions such as the GEVD, the mean and standard deviation are no longer reasonable proxies for the location and scale of the distribution.
Moreover, for {large enough} values of the shape parameter, the GEV mean and variance grow to infinity, which prevents us from interpreting the location-scale GEV parametrisation as we {usually} do in other models.
This is particularly troublesome in a regression context where the natural choice is to use a regression model in the location parameter, the scale parameter, or both.
In such cases, covariate coefficients may not have a clear meaning if the GEV mean and variance are not defined. 
This is even more critical in a Bayesian context, as it might not be clear how to assign reasonable priors to GEV parameters. 
In this section, we propose a reparametrisation of the GEV that has a broader interpretation for all values of $\xi\in\R$, where prior information can be assigned to parameters that can still be widely interpreted even in the cases where the first and second moments are not finite. 
Specifically, following the spirit of \cite{coles1996bayesian}, we reparametrise the GEVD in terms of new quantile-based location and spread parameters.  
The new location parameter is the $\alpha$-quantile ($0< \alpha< 1$), denoted $q_\alpha\in\R$, while the new spread is the quantile range $s_\beta = q_{1-\beta/2} - q_{\beta/2}\in\R^+$ $(0< \beta < 1$).
So, for instance, if $\alpha = \beta = 0.5$, then the new location $q_{0.5}$ is the median, and the spread $s_{0.5}$ is the interquartile range.
The GEVD reparametrised in terms of $q_\alpha,s_\beta$ and $\xi$ can be written as
$$F(y\mid q_\alpha,s_\beta,\xi) = \exp\left[-\left\{\frac{y-q_\alpha}{s_\beta(\ell_{1-\beta/2,\xi} - \ell_{\beta/2,\xi})^{-1}} + \ell_{\alpha,\xi}\right\}_+^{-1/\xi}\right],$$
where $a_+=\max(a,0)$ and for any $a>0$, $\ell_{a,\xi} = (-\log a)^{-\xi}$. Note that the case $\xi=0$ simplifies to
$$F(y\mid q_\alpha,s_\beta) = \exp\left[-\exp\left\{-\left(\frac{y - q_\alpha}{s_\beta(\ell_{\beta/2} - \ell_{1-\beta/2})^{-1}} - \ell_\alpha\right)\right\}\right],$$
with $\ell_a = \log(-\log a)${, for any $a>0$.} 
There is a one-to-one mapping between $(q_\alpha, s_\beta, \xi)^\top$ and the parametrisation {in terms of $(\mu,\sigma,\xi)^\top$}.
For the case $\xi\neq 0$, the mapping is given by
$$\mu=q_\alpha-\frac{s_\beta(\ell_{\alpha,\xi}-1)}{\ell_{1-\beta/2,\xi} - \ell_{\beta/2,\xi}}, \qquad \sigma = \frac{\xi s_\beta}{\ell_{1-\beta/2,\xi} - \ell_{\beta/2,\xi}}.$$
The case $\xi=0$ is interpreted as the limit when $\xi\to0$, i.e.,
$$\mu = q_\alpha + \frac{s_\beta\ell_\alpha}{\ell_{\beta/2} - \ell_{1-\beta/2}},\qquad \sigma = \frac{s_\beta}{\ell_{\beta/2} - \ell_{1-/\beta/2}}.$$

Note that the location and spread parameters are still interpretable in this new parametrisation even when the mean and variance are undefined.
The probabilities $\alpha$ and $\beta$ that define the quantiles $q_\alpha,q_{\beta/2}, q_{1-\beta/2}$ are fixed hyperparameters and should be defined by the user.

\section{The blended GEV model}~\label{sec:bGEV}
As mentioned in Section~\ref{sec:introduction}, we want to modify the left tail of the GEVD with positive shape parameter to avoid inheriting a left-bounded support {for} the finite-sample maxima.
We want to do this by blending two distributions, {$F(x)$ and $G(x)$}, at the distribution level, so that the resulting distribution {$H(x)$ resembles the GEVD's} right tail, has infinite support, and the resulting density function can straightforwardly be computed and is automatically normalised.
Our blended GEV (bGEV) distribution {function is defined as}
\begin{equation}\label{eq:bGEV}
H(x \mid q_{\alpha},s_{\beta}, \xi) = F(x \mid q_{\alpha}, s_{\beta}, \xi)^{p(x)} G(x \mid \tilde{q}_{\alpha}, \tilde{s}_{\beta})^{1-p(x)},
\end{equation}
where $F$ is the {Fr\'echet} distribution with location $q_{\alpha}$, spread $s_{\beta}$, and shape $\xi$, and $G$ is the Gumbel distribution with location $\tilde{q}_{\alpha} \equiv \tilde{q}_{\alpha}(q_{\alpha},s_{\beta})$ and spread  $\tilde{s}_{\beta} \equiv \tilde{s}_{\beta}(q_{\alpha},s_{\beta})$.
The function $p$ is a weight function that controls the way the distributions $F$ and $G$ are blended together.

\begin{figure}[!htbp]
	\centering
	\includegraphics[scale = 0.35]{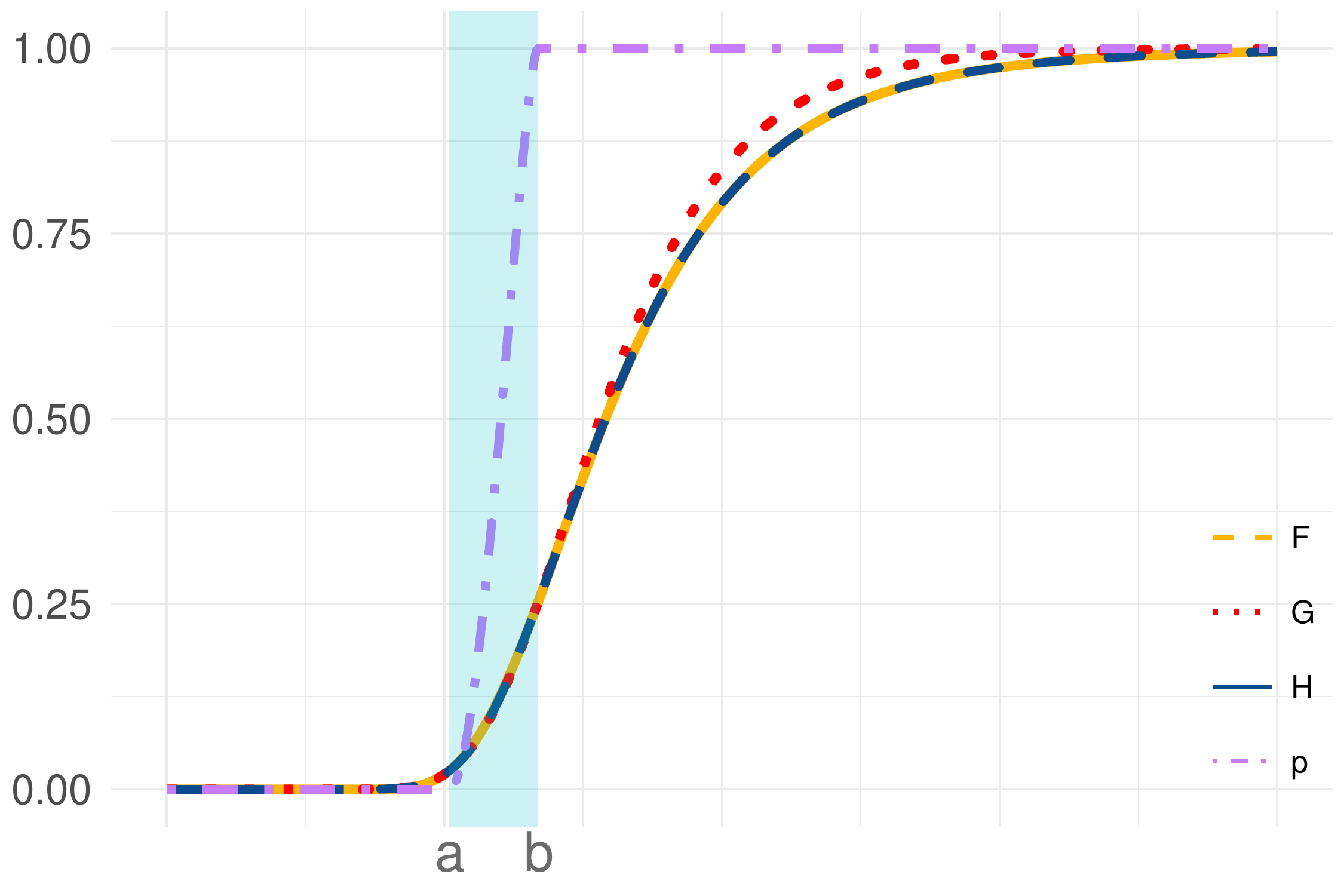}
	\includegraphics[scale = 0.35]{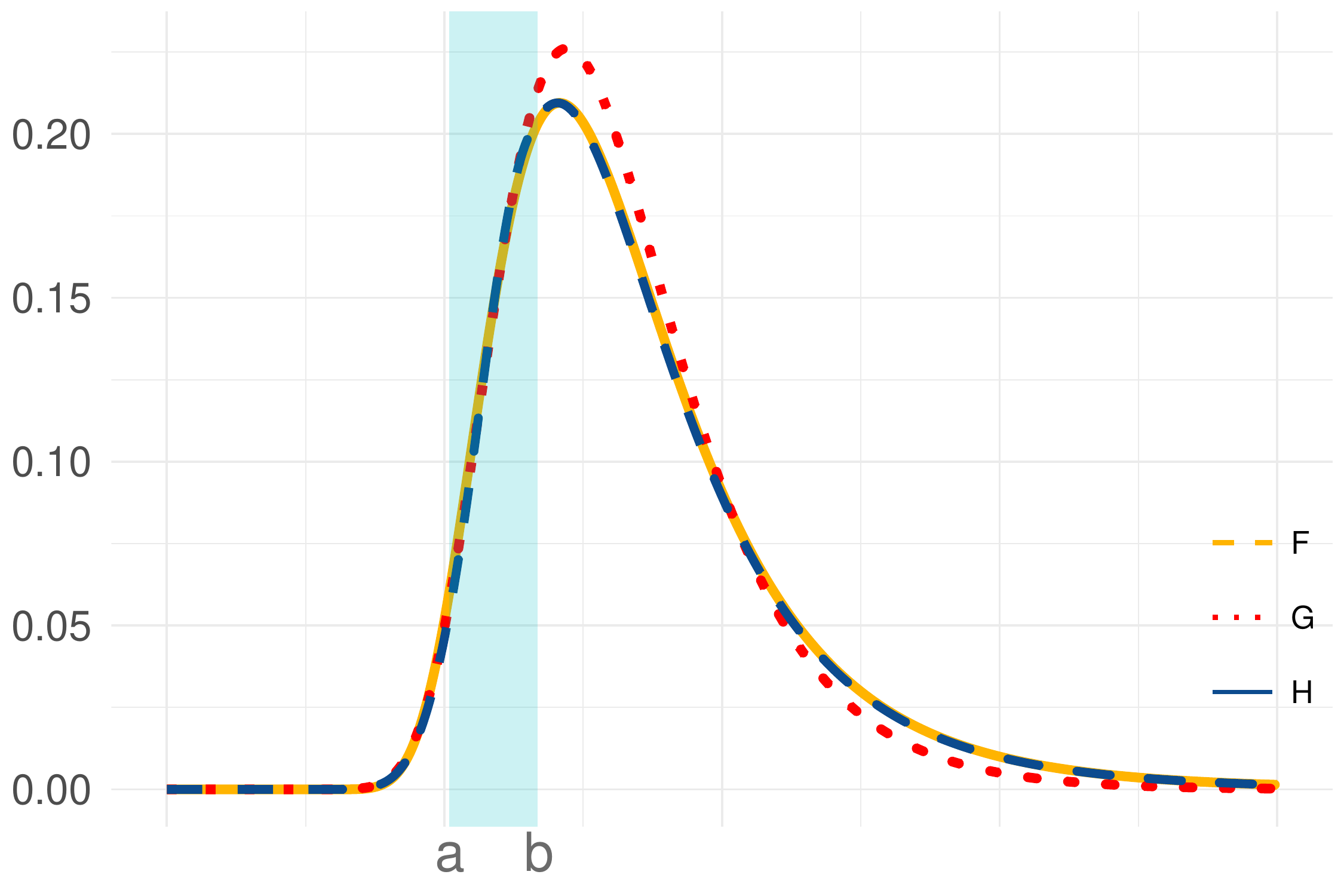}
	\caption{\footnotesize Illustration of the distribution (left) and density (right) for $H$ (bGEV), $F$ ({Fr\'echet}), and $G$ (Gumbel). On the left, the purple dash-dotted line is the weight function $p$.}
	\label{fig:distden.pdf}
\end{figure}

As per the notation in \eqref{eq:bGEV}, $H$ is parametrised solely in terms of $q_{\alpha},s_{\beta}$ and $\xi$, and not in terms of $\tilde{q}_{\alpha}$ and $\tilde{s}_{\beta}$.
Indeed, in the following we will see that $\tilde{q}_{\alpha}$ and $\tilde{s}_{\beta}$ can be completely characterised via $q_{\alpha},s_{\beta},\xi$ and a set of hyperparameters, of which the first two are the probabilities $\alpha$ and $\beta$.
In principle, $\alpha$ and $\beta$ can take any value, but as noted in Section~\ref{sec:reparam}, some choices might be easier to interpret than other (e.g., setting $\alpha=\beta=0.5$).
Three other hyperparameters need to be defined, and they are all related to the weight function $p$.
The function $p$ is chosen as the cumulative distribution function of a beta distribution with shape parameters $c_1>0$ and $c_2>0$ ($F_{\text{Beta}}(\cdot\mid c_1,c_2)$) defined over the \emph{left-tail mixing area} $[a,b]$.
As the name implies, the left-tail mixing area is where $F$ and $G$ are merged; see the blue shaded rectangle in Figure~\ref{fig:distden.pdf}.
The weight function $p$ is then defined as $F_{\text{Beta}}(\cdot\mid c_1,c_2)$  evaluated {at} the point $(x - a)/(b - a)$, i.e.,
\begin{equation}\label{eq:weightfunction}
	p(x) = p(x ;a, b,c_1,c_2) = F_{\text{Beta}}\left(\frac{x-a}{b-a} \mid c_1, c_2\right).
\end{equation}

The left-tail mixing area $[a,b]$ is defined by setting $a = F^{-1}(p_a)$ and $b = F^{-1}(p_b)$ for {relatively} small probabilities $p_a$ and $p_b$. 
The shape parameters $c_1$ and $c_2$ of the weight function $p$ can take any positive value, but we restrict them to greater than 3. 
In that case, the second derivative of the log-density of $H$ is always continuous (see Section~\ref{ap:derivbgevdensity} in the Appendix).

Some comments related to our above choices are in order.
First, note that the weight function $p$ in \eqref{eq:weightfunction} is 0 for $x\leq a$ and one for $x\geq b$, which means that the left tail of $H$ is equal to the left tail of $G$ (which is unbounded) and the right tail of $H$ is equal to the right tail of $F$.
In other words, the distribution $H$ has the correct limiting right tail, {yet without the inconvenient, left-bounded, parameter-dependent support} imposed by {the} limit.
Note also that $\lim_{\xi\to 0} H(x) = G(x)$ for all {weight functions} $p$, which means that our proposed distribution reduces to the {classical} Gumbel case when $\xi\to0$, for all $p$. 
{This also ensures that the resulting distribution $H$ is not too far from the classical GEVD when $\xi\approx 0$.}
Second, note that by setting $a = F^{-1}(p_a)$ and $b = F^{-1}(p_b)$ we get that $F(a) = G(a)$, $F(b) = G(b)$ and the resulting distribution $H$ is continuous.
Moreover, we can express the Gumbel parameters $\tilde{q}_{\alpha}$ and $\tilde{s}_{\beta}$ as
\begin{equation}\label{eq:ab}
	\tilde{q}_{\alpha} = a - \frac{(b-a)(\ell_{\alpha} - \ell_{p_a})}{\ell_{p_a} - \ell_{p_b}},\qquad 
	\tilde{s}_{\beta} = \frac{(b-a)(\ell_{\beta/2}-\ell_{1-\beta/2})}{\ell_{p_a}-\ell_{p_b}},
    \end{equation}
which are injective functions of $q_{\alpha}$ and $s_{\beta}$.
Finally, note that the reparametrisation proposed in Section~\ref{sec:reparam} is based on the quantiles $q_\alpha$, $q_{\beta/2}$ and $q_{1-\beta/2}$ of the GEVD.
Since the bGEVD is equal to the GEVD only for values above the left-tail mixing area (i.e., only for $x\geq b$), the definition of $q_\alpha$ and $s_\beta$ should be constrained to the value of $b$.
Since $b$ is also defined as a quantile of the GEVD, the above implies that we should have $\alpha\geq p_b$ and $\beta/2\geq p_b$.
Otherwise, $q_\alpha$ and $s_\beta$ would be defined as quantiles of the bGEVD, which are different from those of the GEVD below $b$.

Summarising, $H$ is completely defined through the parameters $(q_{\alpha},s_{\beta},\xi)$ and the hyperparameters $(\alpha,\beta,p_a,p_b, c_1,c_2)$.
For our data application in Section~\ref{sec:application}, we set $\alpha=\beta = 0.5$, in which case $H$ is parametrised in terms of the median and the interquartile range of $F$.
Additionally, we set $p_a=0.05$, $p_b=0.2$ and $c_1=c_2=5$.
The choice for $c_1$ and $c_2$ leads to a symmetric and computationally convenient weight function.

\section{Property-preserving penalised complexity priors}\label{sec:pppc}

Penalised complexity (PC) priors~\citep{simpson2017penalising} provide a principled and widely applicable method to specify prior information that is difficult to obtain from expert knowledge.
It uses the natural nested structure inherent in many model components to define these model components as flexible extensions of some base model. 
The method penalises deviations from this base model by placing an exponential prior on the Kullback-Leibler  {divergence} from the base model.
Base models should then be defined for every model component of interest.
This task is simplified by noticing that, in practice, model components are completely defined through parameters.
Take, for example, a Gaussian random walk of order 2 (a definition can be found in~\eqref{eq:rw}), which is entirely defined by its precision (inverse of its standard deviation).
This fact can be used to define a PC prior using as base model a random walk with a infinite precision (see Section~\ref{sec:application} for an application of this particular PC prior).

PC priors can also be defined over data parameters, such as the positive shape parameter of the GEV/bGEV distributions.
The natural base model, in this case, is a GEV/bGEV with a shape equal to 0 (i.e., a Gumbel distribution).
Although it is possible to derive a formula for this PC prior, we can take advantage of the relationship between the GEV and generalised Pareto (GP) distributions and approximate the PC prior of the GEV shape by that of the GP shape.
\cite{opitz2018inla} derived the PC prior for the GP positive shape parameter using a GP with shape equal to 0 as base model.
This PC prior is defined for $0\leq \xi<1${, thus preventing infinite-mean models,} and depends on a penalty parameter $\lambda>0$.
It can be expressed as
\begin{equation}\label{eq:pcgp}
    \pi(\xi) = \frac{\lambda}{\sqrt{2}}\exp\left(-\frac{\lambda}{\sqrt{2}}\frac{\xi}{(1-\xi)^{1/2}}\right)\left(\frac{1-\xi/2}{(1-\xi)^{3/2}}\right){,\qquad \xi\in[0,1).}
\end{equation}
We can use the PC prior framework to define priors that help preserve specific properties of the model components or the data distribution, such as the existence of moments.
This is what we call the property-preserving penalised complexity (P$^3$C) prior approach.
In the case of the GEVD, the existence of the $k$-th moment is guaranteed if  $\xi<1/k$.
{Although,} in practice, the existence of high-order moments for the GEVD with $\xi$ strictly positive cannot be guaranteed, 
we can preserve low-order moments by shrinking the interval of possible values that $\xi$ can take \emph{a priori}.
Specifically, we can preserve the first two moments of the GEVD by conditioning on $\xi$ being less than 0.5.
In practice, this means that we normalise~\eqref{eq:pcgp} by the corresponding cumulative distribution evaluated at 0.5, i.e., using the alternative prior
\begin{equation}\label{eq:pcgp2}
\tilde{\pi}(\xi)={\pi(\xi)\over\int_0^{0.5}\pi(s){\rm d}s},\qquad \xi\in[0,0.5).
\end{equation}
A method such as the one presented here is particularly important in prediction problems.
Indeed, if the prior for $\xi$ has positive support in $[0,u_\xi]$, then for finite sample sizes, the prediction of a new data point will inherit the moment properties of $\xi\in[0,u_\xi]$.
Therefore, if, for instance, $u_\xi=1$, the posterior distribution will assign a positive density to $\xi\in[0,1]$, which means that all predictions for a new data point will have infinite mean and variance once the uncertainty is taken into account, even if the data suggest the opposite.

The P$^3$C prior concept that we show here for the shape parameter can be extended to other models' parameters when important model properties are not \emph{continuous} as a function of such parameters.
We argue that distributional properties such as the existence of moments are too crucial to be {determined} by the randomness of the data-generating process.
Therefore, the importance of the P$^3$C prior approach lies in providing a framework where we can make rational decisions about important model properties through prior knowledge, avoiding any influence of the data or the model error inherent to any statistical model fit.

Note that, in principle, standard bounded priors over $\xi$ (such as a Beta$(0,1)$ distribution
conditioned to be less than $1/2$) can also be used to preserve the first two moments. 
Here we choose the PC prior approach because it is a principled method flexible enough to represent different levels of prior knowledge, but alternative approaches could also be taken.

\section{Simulation}~\label{sec:simulation}
We conduct simulation studies to compare the performance of the GEVD and the bGEVD at estimating return levels and to provide some insights on the effects of the hyperparameters $\alpha,\beta,p_a, p_b$ and $c_1=c_2$ on the bGEV model.
Since the GEVD is the only possible non-degenerate limit to properly renormalised maxima, we expect the GEVD to outperform the bGEV for large sample sizes.
However, it is very rare to find a sufficiently large number of block-maxima in practical applications.
In those cases, we expect both models to perform similarly.
For our first simulation study, we draw $n$ samples from a Fr\'echet distribution with location 0, scale 1 and shape $\alpha=1/0.1$ and retain the largest value. 
We repeat this $N$ times to get a sample of $N$ independent block-maxima and fit the GEV and bGEV distributions using maximum likelihood.
We then use the fitted values to compute return levels associated with a return period $T$ and compare them with the return levels from the true maxima distribution (Fr\'echet$(0,n^{1/\alpha}, \alpha)$).
We repeat the above $M$ times to obtain Monte Carlo performance measures.
Table~\ref{table:sim1} show our results in terms of the difference $(\text{RMSE}_{\text{GEV}}-\text{RMSE}_{\text{bGEV}})$, where $\text{RMSE}_{\text{GEV}}$ and $\text{RMSE}_{\text{bGEV}}$ are the root mean square error associated to the GEV and bGEV fits, respectively.
These measures where computed based on $M=500$ Monte Carlo replicates with $n=30,50,100, 500$, $N=30,50,100,500, 1000$ and $T=30,50,100$.
We can see that both models perform similarly for all block sizes ($n$), number of block-maxima ($N$) and return periods ($T$).
As expected, the GEVD performs slightly better for large number of block-maxima (see the last column, where $N=1000$).
We can also see that for $N=30$, the bGEVD performs slightly better than the GEVD for most block sizes.
Note that both models were fitted using the same optimisation method~\citep{nelder1965simplex} and the same starting values.

\begin{table}[!htbp]
\centering
\begin{tabular}{c | c |c |c |c |c}
  \toprule
 $N$ & 30 & 50 & 100 & 500 & 1000\\ 
 $n$&{$T=30/50/100$} & {$T=30/50/100$} & {$T=30/50/100$} & {$T=30/50/100$} & {$T=30/50/100$}\\
  \hline
30 & 0/0/0.01 & 0/-0.01/-0.01 & 0/0/0 & 0/-0.01/0 & 0/0/-0.01 \\ 
  50 & 0/0/0.01 & 0/0/0 & 0/0/0 & 0/0/0 & 0/0/-0.01 \\ 
  100 & -0.01/0/0 & 0/-0.01/0 & 0/0/-0.01 & 0/0/0 & 0/0/-0.01 \\ 
  500 & 0/0/0.01 & 0/0/0 & 0/0/0 & 0/0/-0.01 & -0.01/0/-0.01 \\ 
  \bottomrule
\end{tabular}
\caption{Difference $(\text{RMSE}_{\text{GEV}}-\text{RMSE}_{\text{bGEV}})$ for return levels associated with different return periods $T$ based on $M=500$ Monte Carlo replicates. $N$ is the number of block-maxima and $n$ is the block size.}
\label{table:sim1}
\end{table}

Our second simulation study aims to assess the effect of the hyperparameters $p_a, p_b$ and $c_1=c_2$ over the bGEV model.
To this end, we generate $N=100$ samples from the GEV distribution with $\mu=0$, $\sigma=1$ and $\xi=0.1$.
We then fit the bGEV using maximum likelihood and different values of the hyperparameters $p_a, p_b$ and $c_1=c_2$.
Specifically, the values are $p_a=0.05,0.1,0.15$, $p_b=0.2,0.25,0.3$ and $c_1=c_2=3,5$. 
We compare the different models based on their ability to estimate a theoretical 50-year return level and summarise our results by computing the RMSE based on $M=500$ Monte Carlo replicates.
Our results in Table~\ref{table:sim2} show no major differences across all 18 configurations of $p_a,p_b$ and $c_1=c_2$, although for fixed values of $p_b$, there seems to be a slight increase in the RMSE as $p_a$ increases, i.e., as the left-tail mixing area become smaller.
Note that the left-tail mixing region, defined by the $p_a$- and $p_b$-quantiles of the Frech\'et distribution $F$, conditions the choice for $\alpha$ and $\beta$.
Indeed, as mentioned in Section~\ref{sec:bGEV}, we must have $p_a<p_b\leq\min\{\alpha,\beta/2\}$.
Therefore, it is advisable to choose $p_a$ and $p_b$ to be relatively small probabilities.
In light of the results from Tablee~\ref{table:sim2}, it is also advisable to choose them not too close.
Finally, note that all the bGEV RMSEs are better than the RMSE obtain with GEV fit, which is equal to 2.53.
\begin{table}
\centering
  \begin{tabular}{l |S S |S S |S S}
    \toprule
    $p_a$ &
      \multicolumn{2}{c}{$0.05$} &
      \multicolumn{2}{c}{$0.1$} &
      \multicolumn{2}{c}{$0.15$} \\
      \hline
      $p_b$& {$c_1=c_2=3$} & {$c_1=c_2=5$} & {$c_1=c_2=3$} & {$c_1=c_2=5$} & {$c_1=c_2=3$} & {$c_1=c_2=5$} \\
      \hline
0.2 & 1.11 & 1.11 & 1.13 & 1.13 & 1.14 & 1.14 \\ 
  0.25 & 1.11 & 1.11 & 1.13 & 1.13 & 1.14 & 1.14 \\ 
  0.3 & 1.11 & 1.11 & 1.13 & 1.13 & 1.14 & 1.14 \\ 
    \bottomrule
  \end{tabular}
   \caption{RMSE for 50-year return level based on $M=500$ Monte Carlo replicates to assess the effect of the hyperparameters $p_a,p_b$ and $c_1=c_2$ when fitting the bGEV model. For reference, the RMSE associated with a GEV fit is 2.53.}
\label{table:sim2}
\end{table}

Our third simulation study analyses the effect of the hyperparameters $\alpha$ and $\beta$ on the bGEV model.
To this end, we follow the same setup used in our second simulation study with $p_a=0.05, p_b=0.2, c_1=c_2=5$ and $\alpha = (0.3, 0.5, 0.7, 0.9)$ and $\beta = (0.5, 0.7, 0.9)$. 
The spreads $s_\beta$ associated with the different values of $\beta$ are, respectively, the length of the central $50\%, 30\%$ and $10\%$ interval.
As before, we compare the different models based on their ability to estimate a theoretical 50-year return level and summarise our results by computing the RMSE based on $M=500$ Monte Carlo replicates.
From Table~\ref{table:sim3}, we can see that combinations of $\alpha$ and $\beta$ yield different levels of accuracy with no clear monotonic pattern.
Except for two cases ($\alpha=0.7$ and $\beta=0.7,0.9$), the bGEV fit with any combination of $\alpha$ and $\beta$ leads to a better RMSE than the GEV fit, whose RMSE is 2.53.
\begin{table}[!htbp]
\centering
\begin{tabular}{r | rrr}
     \toprule
$\alpha/\beta$ & 0.5 & 0.7 & 0.9 \\ 
  \hline
0.3 & 1.47 & 1.11 & 0.57 \\ 
  0.5 & 1.88 & 1.63 & 0.91 \\ 
  0.7 & 1.92 & 2.17 & 2.16 \\ 
  0.9 & 0.75 & 0.65 & 0.59 \\ 
   \bottomrule
\end{tabular}
\caption{RMSE for 50-year return level based on $M=500$ Monte Carlo replicates to assess the effect of the hyperparameters $\alpha$ and $\beta$ when fitting the bGEV model. For reference, the RMSE associated with a GEV fit is 2.53.}
\label{table:sim3}
\end{table}

To summarise, our first simulation study shows that the GEV and bGEV are comparable when estimating the right tail.
Note that in this simulation study, maxima were taken over Fr\'echet random variables with location parameter equal to 0, which means that the generated values are bounded by 0.
This represents a setting where the original observations have a natural (physical) lower bound, which is the case for, e.g., precipitation or pollutant concentrations.
From our results in Table~\ref{table:sim1}, we can see that the bGEV model misspecification (support in the whole real line) does not affect the resulting upper tail estimation.
Finally, the results presented in Tables~\ref{table:sim2} and \ref{table:sim3} show the effect of the hyperparameters associated with the new parametrisation, the left-tail mixing region and the weight function and can be used as guidelines when fitting the bGEVD.

\section{Application}~\label{sec:application}
In this section, we illustrate the bGEV model using the new parametrisation (Section~\ref{sec:reparam}) and the the P$^3$C prior approach (Section~\ref{sec:pppc}).
To this end, we consider monthly {maximum} concentrations of nitrogen dioxide (NO$_2$), measured in microgrammes per cubic meter or $\mu g/m^3$, in Bakersfield, a city in Kern County, California.
NO$_2$ is a chemical compound that primarily gets in the air from fuel-burning (U.S. Environmental Protection Agency, \url{epa.gov}); see \citet{vettori2019bayesian,Vettori.etal:2020} for related studies of extreme air pollutant concentrations in California.
Monthly maximum NO$_2$ measurements are computed from January 2000 to March 2016, giving rise to 136 complete measurements.
Available covariates include monthly mean and monthly maximum measurements of wind speed, temperature, pressure, and relative humidity (eight covariates in total). 
To model (possible non-linear) relationships between the covariates and the NO$_2$ concentrations, we assume that monthly {maximum} concentrations of NO$_2$ follow a bGEVD where the location $q_\alpha$ and spread $s_\beta$ vary {with covariates}. 
For simplicity, we assume the shape $\xi$ to be covariate-free.
Although likelihood-based inference assuming independence is straightforward (see the density associated with the model in~\eqref{eq:bGEV} in the Section~\ref{ap:bgevdensity} of the Appendix), we here consider Bayesian inference in the context of latent Gaussian models using the integrated nested Laplace approximation (INLA; \citealp{rue2009approximate}).
Within the INLA framework, we assume that observations are conditionally independent given a latent Gaussian field and a set of hyperparameters.
Loosely speaking, the latent field describes the underlying trends and dependence structure of the observations and can be represented in terms of an additive linear predictor.
For the bGEV model, we link the linear predictor to $q_\alpha$, which means that the linear predictor describes the effect of the covariates over the $\alpha$-quantile.
Additionally, our bGEV-INLA implementation allows the spread and the shape parameters to vary linearly with covariates.

The goal of the Bayesian inference is to get posterior marginal distributions for the components of the linear predictor, the components of the linear models for the spread and shape parameters, and the set of hyperparameters.
This translates into the computation of high-dimensional integrals, and INLA tackles these computations conveniently using the Laplace approximation in a nested way, leveraging modern numerical techniques for sparse matrices.
It is important to notice that latent Gaussian models represent a very vast class of statistical models; therefore, INLA can be used in a variety of applications (see, e.g., \citealp{wang2018bayesian} and \citealp{krainski2018advanced}). 
For a gentle review of the INLA methodology, see~\citealp{rue2017bayesian}.

Model selection was carried out using the Watanabe–Akaike Information Criterion (WAIC), the Deviance Information Criterion (DIC), and graphical tools based on the probability integral transform over all the possible configurations of the eight covariates when we include them linearly and non-linearly in the location $q_\alpha$ and linearly in the spread $s_\beta$.
The best model obtained using a forward-selection procedure has a covariate-dependent location and covariate-free spread and shape and is given by
\begin{align}\label{eq:inla}
    q_{\alpha}(t) &= \gamma_0 + \gamma_1\mathbbm{1}_{\texttt{yot}_t=i} + \gamma_2\times\texttt{monthly max wind speed}(t) +{f_{1}}(\texttt{month}({t})) \nonumber\\
    &+f_{2}(\texttt{monthly mean temperature}({t})),\quad i = 1,\ldots,17,\nonumber\\
    \log(s_{\beta}(t)) &= s_{\beta,0},\qquad \xi(t) = \xi_0,
\end{align}
for $t=1,\ldots,136$, where $\mathbbm{1}_{\texttt{yot}_t=i}$ denotes the year of time $t$, $\gamma_0$ is an intercept, $\gamma_1$ and $\gamma_2$ are linear regression coefficient and 
${f_{1}}(\cdot)$ and ${f_{2}}(\cdot)$ are non-linear effects defined as cyclic Gaussian random walks of order two with a sum-to-zero constraint for identifiability purposes. 
A cyclic Gaussian random walk of second order $f$ can be defined as follows: let $\ww = (w_{1},\ldots,w_{K})^T$ be a discretisation of the covariate $\xx$ into $K$ bins. 
Then,
\begin{align}\label{eq:rw}
    f(w_{2})-2f(w_{1}) + f(w_{K})&\sim\mathcal{N}(0, \tau^{-1}),\nonumber\\
    f(w_{j+1})-2f(w_{j}) + f(w_{j-1})&\sim\mathcal{N}(0, \tau^{-1}),\quad j = 2,\ldots,K-1,
\end{align}
is a Gaussian random walk of order 2 with precision parameter $\tau$ that controls the level of smoothness of the random walk.
Note that $f_1$ is defined over the original covariate (month) as it is naturally binned, while ${f_{2}}$ is defined over a binned version of monthly mean temperature.

Prior distributions need to be specified for the three linear regression coefficient in~\eqref{eq:inla}, the precision parameters of the two random walks ($\tau_{1}$ and $\tau_2$ for $f_1$ and $f_2$, respectively), $s_{\beta,0}$ and $\xi_0$.
We choose a vague Gaussian prior with zero mean and precision $3\times 10^{-3}$ for the linear regression coefficient ($\gamma_0,\gamma_1,\gamma_2$). 
Preliminary exploratory analysis (not shown) suggests that the values of NO$_2$ vary smoothly with moths, exhibiting a concave relationship.
Using a grid search, $\tau_1$ was kept fixed to a value that reflects this relationship.
For the precision parameter $\tau_2$, we set weak prior distributions where the probability that the standard deviation $1/\tau_2$ is greater than the empirical standard deviation of the response is 0.01.
For the spread, we assume that $s_{\beta,0}\sim\text{Gamma}(3,3)$ a prior. 
Finally, for $\xi$, we use the P$^3$C prior approach in~\eqref{eq:pcgp2} to preserve the existence of the first two moments of the bGEVD, with $\lambda = 7$.

\begin{figure}[!htbp]
	\centering
    \footnotesize \rotatebox{90}{\textbf{\hspace{2.6cm}{\color{white}{aa}}bGEV}}
    \hspace{.2cm}
	\includegraphics[scale = 0.2]{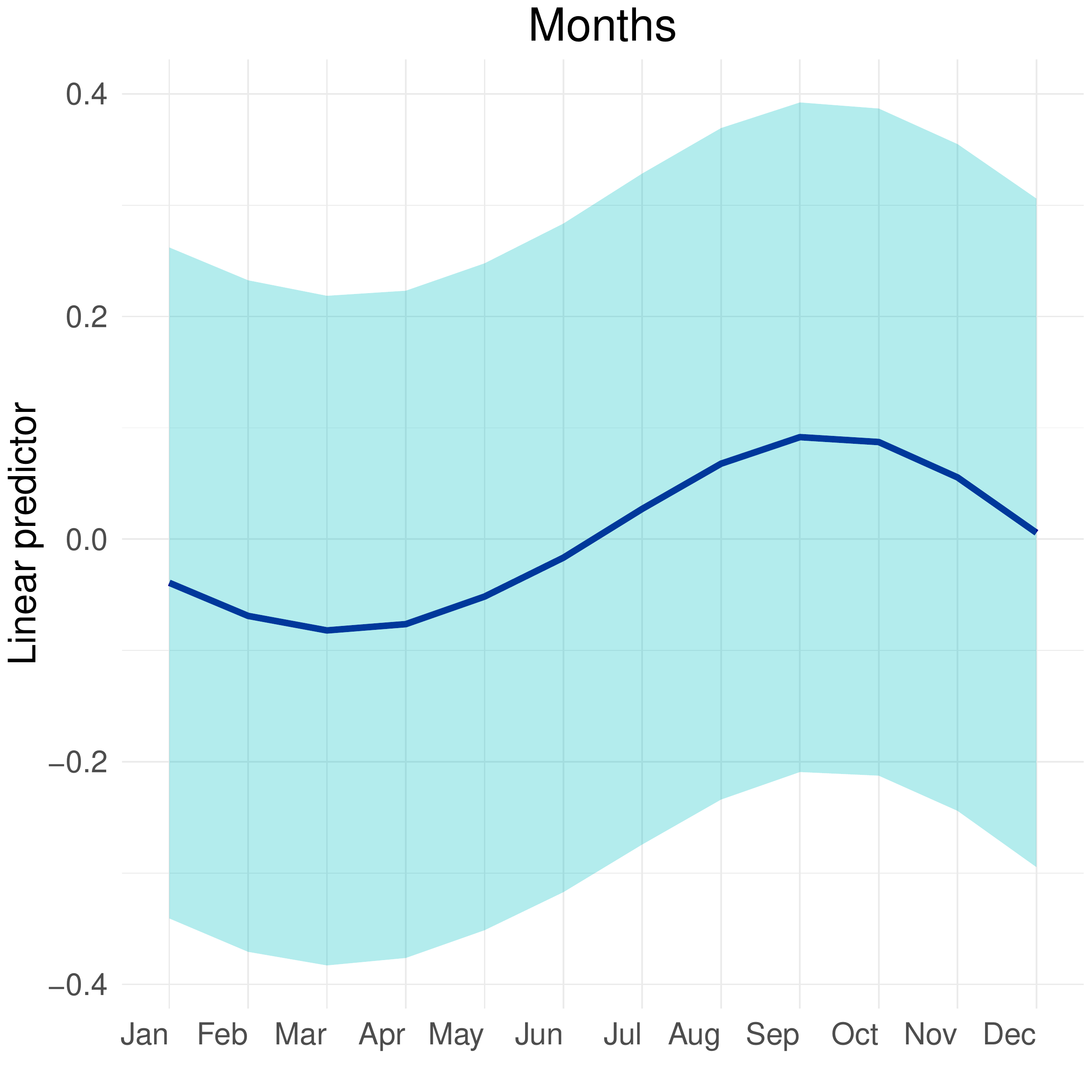}
	\includegraphics[scale = 0.2]{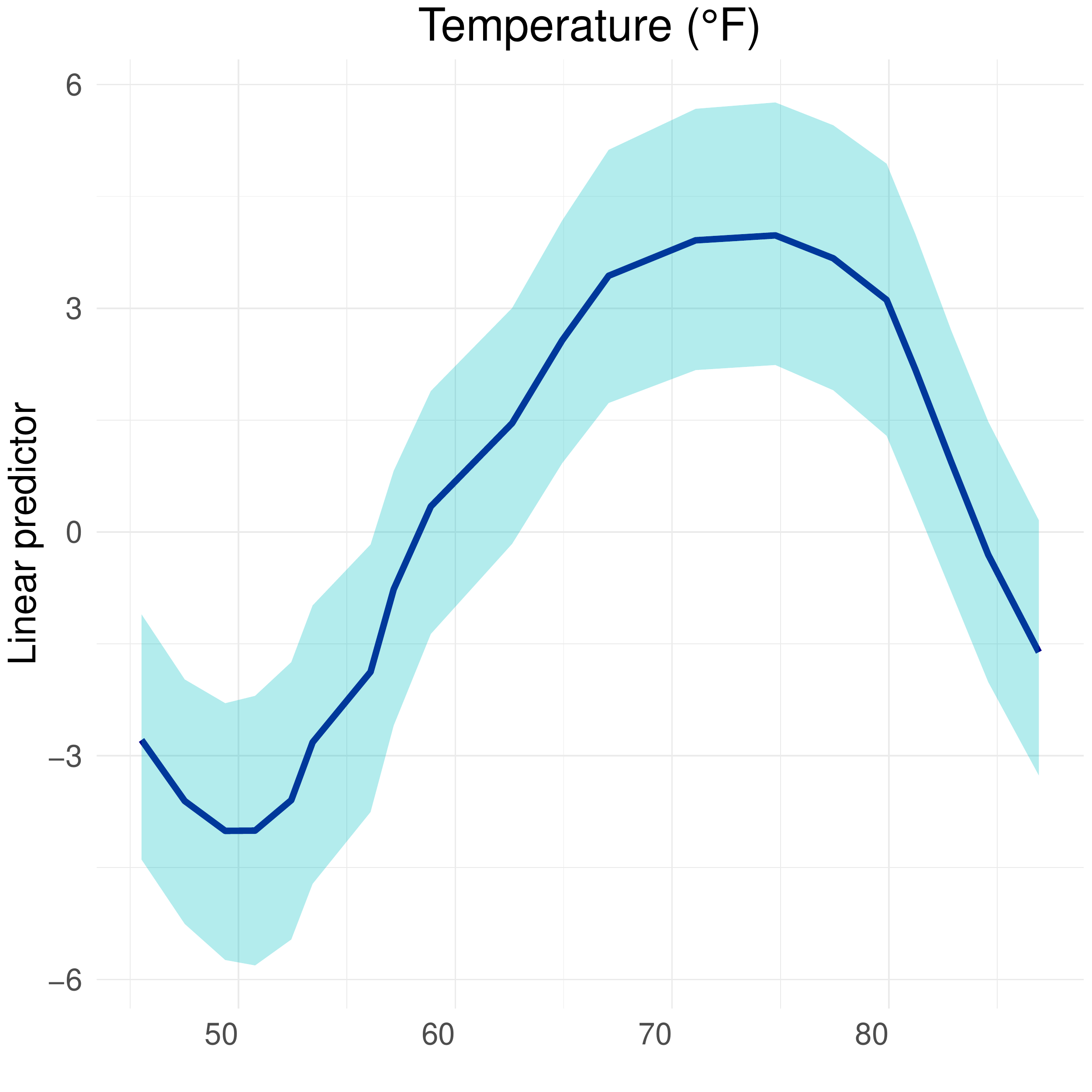}
	\includegraphics[scale = 0.2]{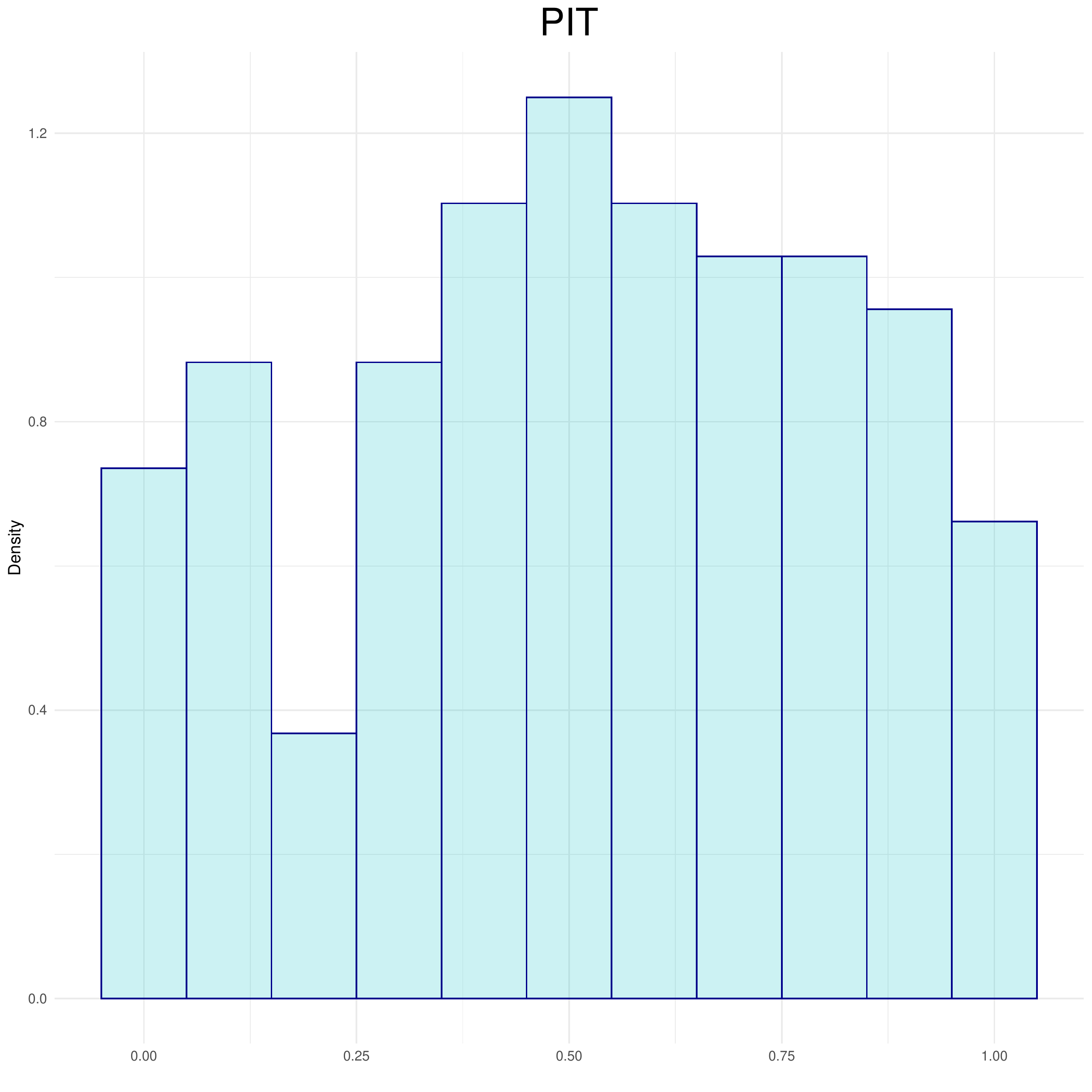}\\
	\footnotesize \rotatebox{90}{\textbf{\hspace{2.8cm}{\color{white}{aa}}GEV}}
    \hspace{.2cm}
	\includegraphics[scale = 0.2]{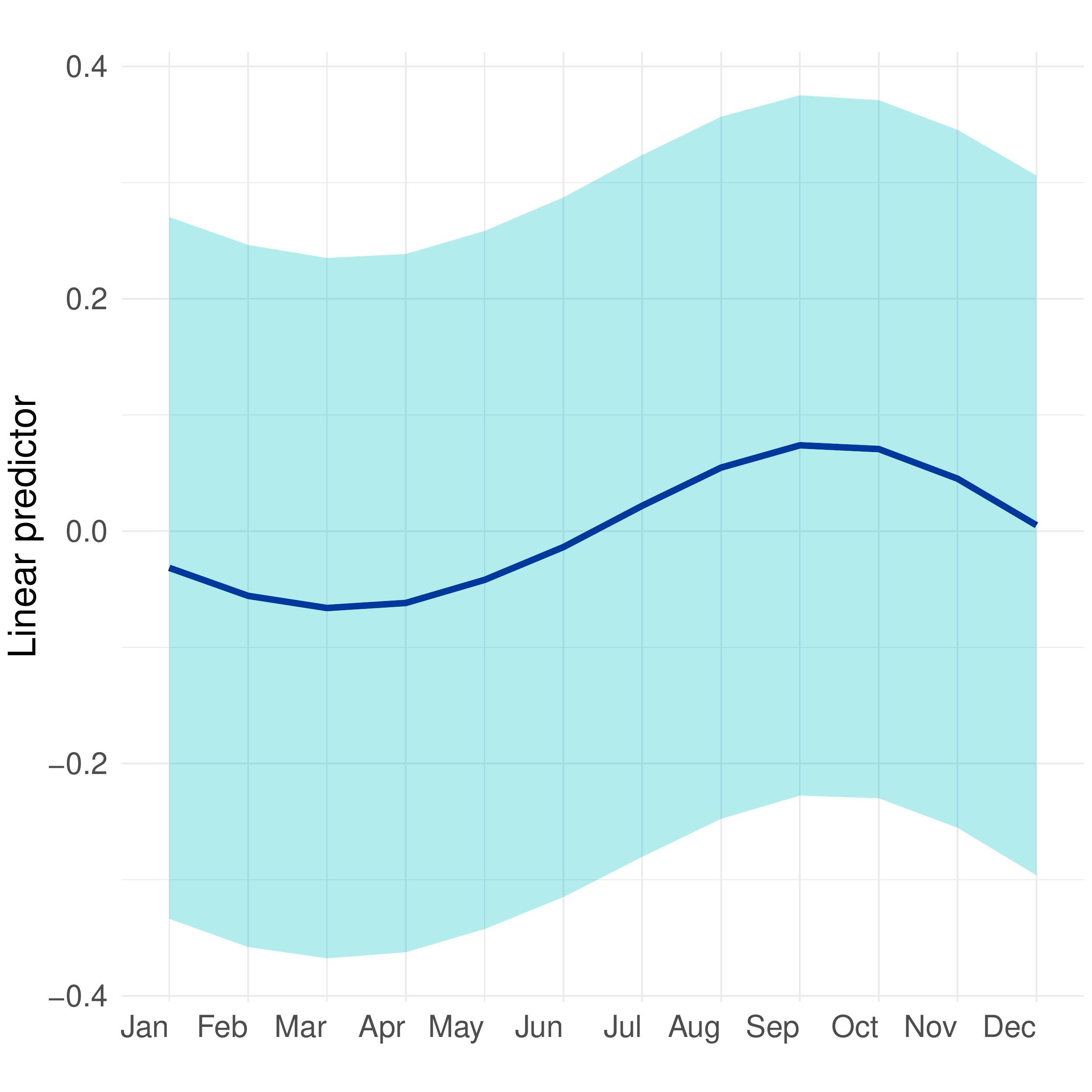}
	\includegraphics[scale = 0.2]{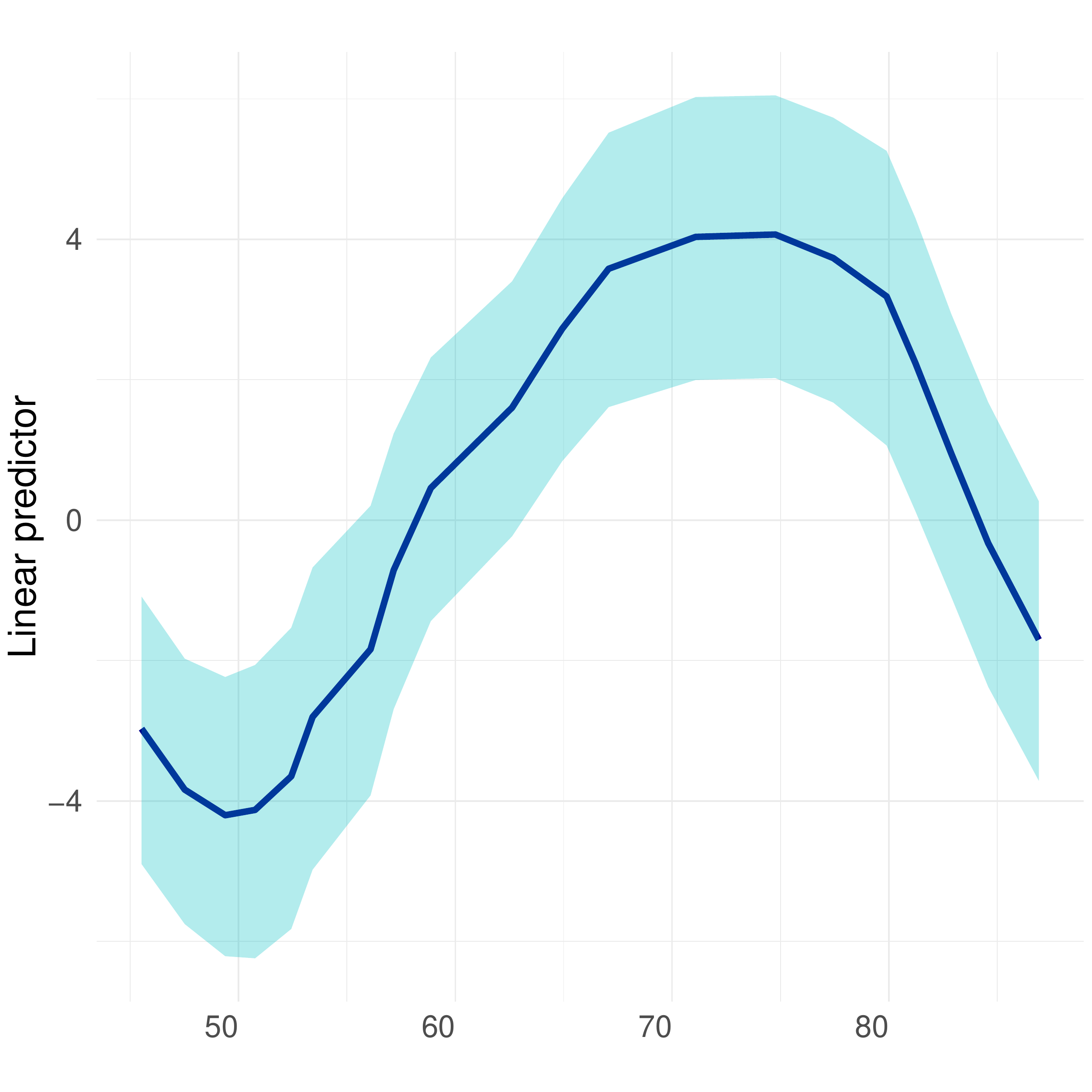}
	\includegraphics[scale = 0.2]{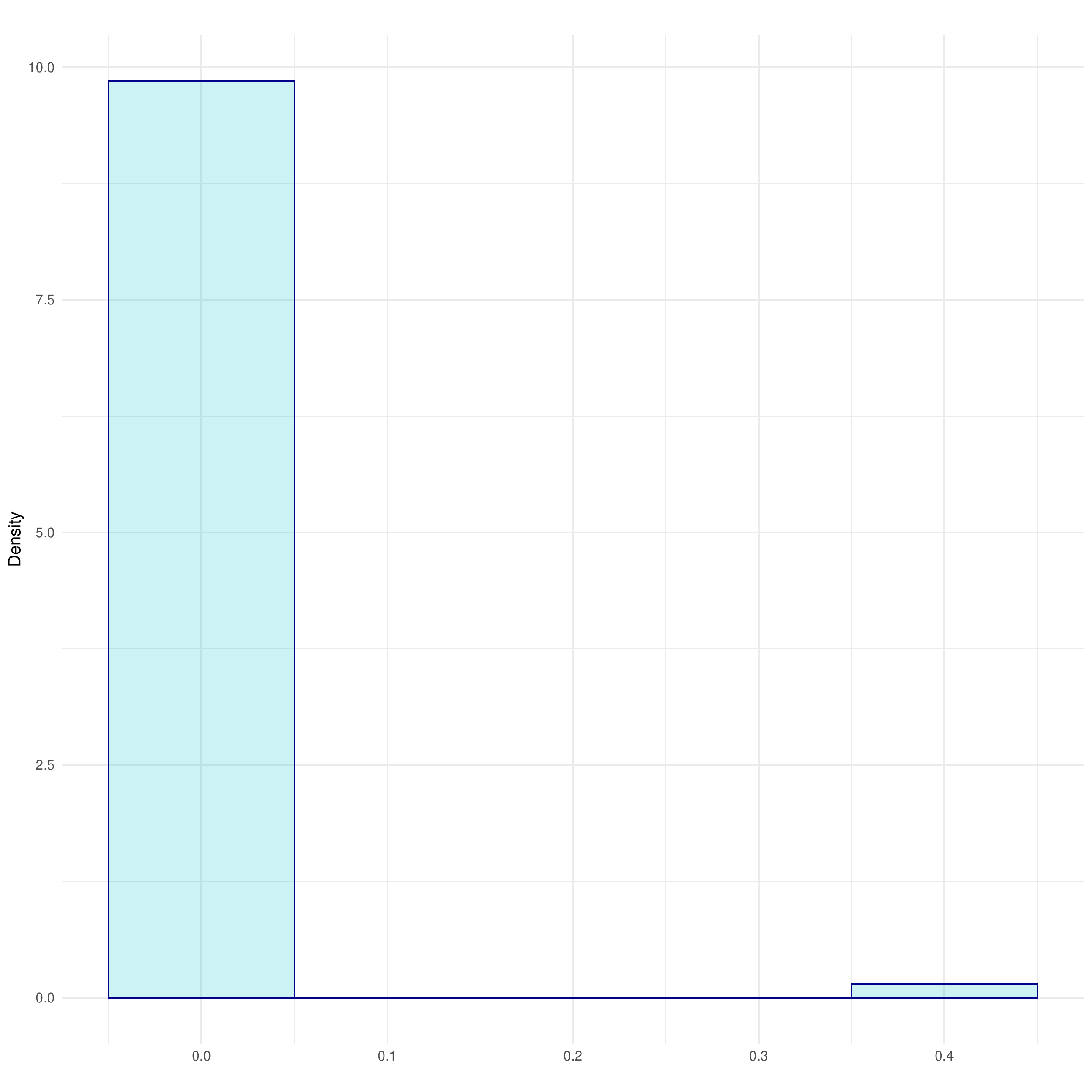}
	\caption{\footnotesize Non-linear effects over months and mean temperature and PIT values for the bGEV (top) and GEV (bottom) models fitted to the NO$_2$ pollution data.}
	\label{fig:app}
\end{figure}

For comparison, we also implement the GEVD with the linear model in~\eqref{eq:inla} linked to the GEV location $\mu(t)$.
Table~\ref{table:app} shows posterior means and 95\% credible intervals for the bGEV and GEV linear regression coefficients and hyperparameters.
Figure~\ref{fig:app} shows the posterior mean and 95\% credible bands of the non-linear effects associated with both fits as well as probability integral transform values (PITs).
PITs are predictive measures of fit, commonly used to assess model calibration~\citep{gneiting2007probabilistic}.
If a model is well-calibrated, the PITs histogram serves as a tool to empirically check for uniformity.
We can see that both fits produce very similar results in terms of model estimates, although the GEV fit had less stable results and required additional steps to guarantee convergence. 
Both models are also comparable in terms of WAIC (934 bGEVD versus 927.8 GEVD) and DIC (938.7 bGEVD versus 926.8 GEVD), but they significantly differ in their prediction ability, as shown by the PITs in Figure~\ref{fig:app}.
Note that the model in~\eqref{eq:inla} was chosen via WAIC and DIC based on the bGEVD, not the GEVD. Therefore, some differences in performance were expected.
However, given that both models produce very similar posterior estimates, the lack of model calibration shown by the GEVD is slightly surprising.

\begin{table}[!htbp]
\centering
\begin{tabular}{rrrr}
  \toprule
 & Pmean & 2.5\% quant & 97.5\% quant \\ 
  \hline
  bGEV intercept & 55.64 & 51.51 & 59.75 \\ 
GEV intercept & 53.80 & 49.19 & 58.32 \\ \hline
bGEV year & -1.14 & -1.37 & -0.92 \\ 
  GEV year & -1.13 & -1.37 & -0.89 \\ \hline
  bGEV max ws & -0.86 & -1.23 & -0.48 \\
  GEV max ws & -0.86 & -1.28 & -0.45 \\ \hline
  bGEV $\tau_2$ & 0.50 & 0.07 & 2.10 \\
  GEV $\tau_2$ & 0.17 & 0.03 & 0.53 \\ \hline
  bGEV $s_\beta$ & 14.33 & 12.65 & 15.94 \\ 
  GEV $\sigma$ & 9.41 & 0.05 & 20.57 \\ \hline
  bGEV $\xi$ & 0.02 & 0.00 & 0.20 \\
  GEV $\xi$ & 0.04 & -0.03 & 0.13 \\ 
   \bottomrule
\end{tabular}
\caption{Posterior means and 95\% credible intervals for the linear regression coefficients and hyperparameters of the bGEV and GEV INLA fits.}
\label{table:app}

\end{table}

\section{Discussion}~\label{sec:discussion}
Inspired by the usually overlooked restrictions inherited by finite-sample maxima distributions when they are approximated by the GEVD, we here make three main contributions.
Firstly, we propose the blended generalised extreme value (bGEV) distribution as an alternative to the {classical} GEVD to {fix} the lower bound constraint imposed by the GEV {distribution} when the shape parameter $\xi$ is positive. 
Specifically, by using the GEV {distribution} as an approximation for the distribution of maxima over blocks, properties of the GEV {distribution}, such as a lower endpoint, are inherited by the finite-sample maxima distribution, which might not be lower-bounded. 
This issue is particularly troublesome in the presence of linear or non-linear covariates and becomes crucial with high-dimensional covariate settings.
The bGEVD {smoothly combines the left tail} of the Gumbel distribution ($\xi=0$) and the right tail of the Fr\'echet distribution ($\xi>0$) and depends on a weight function that controls the influence of both distributions.
It is important to notice that since the GEVD is the only possible non-degenerate limit to properly {renormalised} maxima, we do not propose a different limiting model. 
Instead, we derive a distribution with infinite support that avoids any lower bound restrictions while matching the right tail of the GEVD. 

Secondly, 
we propose a new parametrisation of the GEVD in terms of a quantile and a spread parameter to allow for a more straightforward and meaningful interpretation of the model parameters. 
The {spread} is {here} defined as the difference between two quantiles. 
This parametrisation has significant consequences in a regression setting where parameters are assumed to vary with covariates, as it ensures interpretability even when the first two moments of the distribution are not defined.
The parametrisation has natural connections with empirical quantiles, and it is particularly advantageous in the Bayesian framework as interpretable prior distributions can be easily assigned.

Thirdly, we introduce the {concept} of P$^3$C priors to retain important model properties when these properties are not {``continuous''} as functions of model parameters.
We use this concept to avoid inconsistencies related to the first two moments of the GEVD.
Specifically, since the first two GEV moments are defined for $\xi<0.5$, we restrict the values of $\xi$ to the interval $(0,0.5)$ by normalising the PC prior using the corresponding cumulative distribution.
We argue that the data at hand should not define critical distributional features such as the existence of moments.
The P$^3$C prior approach provides a framework to define those features through prior knowledge.

The GEVD is closely connected to the generalised Pareto distribution (GPD) for threshold exceedances.
Indeed, if the conditions for the extremal types theorem hold, then the GPD is the only possible limiting distribution for exceedances over a threshold $u$ when $u$ grows to infinity.
But unlike the GEVD, the GPD does not have a parameter-dependent support when $\xi>0$, so the methodology presented in Section~\ref{sec:bGEV} is not suitable nor needed.

Note that the new parametrisation introduced in Section~\ref{sec:reparam} is needed when using the bGEV-INLA implementation, but it is not crucial for defining the bGEV model. 
Indeed, we can define the bGEV distribution in terms of the GEV parameters $\mu,\sigma,\xi>0$.
The purpose of the new parametrisation is to provide a way to interpret model parameters even in the absence of the first and second moments, which in turn helps define meaningful priors.

The bGEV model with the new parametrisation and the P$^3$C prior approach is freely available and efficiently implemented in the R-INLA package.
In general, INLA requires the model likelihood to be log-concave~\citep{rue2009approximate}, which is not the case for the GEV and the bGEV distributions.
Although the lack of log-concavity does not necessarily mean that INLA will not converge, it is highly advisable to try to mitigate potential numerical instabilities by, e.g., choosing informative priors for the likelihood parameters.
Additional standardisations of the response variable can also help to reduce convergence issues.

\section*{Notes on implementation}
The INLA implementation and documentation for the bGEV model can be found in the R-INLA package via \texttt{inla.doc("bgev")}. A tutorial for fitting the bGEV model with R-INLA can be found in \url{www.r-inla.org/documentation}. Additionally, codes to reproduce the simulation studies and the data application are freely available from \url{www.github.com/dcastrocamilo/bGEV}.

\section*{Acknowledgements}
 We thank Sabrina Vettori for providing a simplified version of the pollution data. 
We acknowledge Lars Holden for the notion of blending at the distribution level, an idea that came to light during a hallway conversation 20 years ago.
{This publication is partially based upon work supported by the King Abdullah University of Science and Technology (KAUST) Office of Sponsored Research (OSR) under Award No. OSR-CRG2017-3434.}

\baselineskip 16pt
\section*{Appendix}
Here we provide details on the computation of the density of $H$ in~\eqref{eq:bGEV} and its first derivative. For ease of notation, we denote $p(x) = p(x; a,b)$.

\subsection{The bGEV density}\label{ap:bgevdensity}
For $x\in\mathbb{R}$ we can write
\begin{align*}
	H(x) &= \exp\left\{p(x)\log F(x) + (1 - p(x))\log G(x)\right\},\\
	F(x) &= \exp\{-t_1(x)\}, \quad t_1(x) = \{\max(0,z_1(x))\}^{-1/\xi}, \quad z_1(x) = \left\{ \frac{x-q_\alpha}{s_\beta(\ell_{1-\beta/2,\xi} - \ell_{\beta/2,\xi})^{-1}} + \ell_{\alpha,\xi}\right\}_+,\\
	G(x) &= \exp\{-t_2(x)\}, \quad t_2(x) = \exp(-z_2(x)), \quad z_2(x) = \frac{x - \tilde{q}_\alpha}{\tilde{s}_\beta(\ell_{\beta/2} - \ell_{1-\beta/2})^{-1}} - \ell_\alpha.
\end{align*}
The density for $H$, $h$, can then be easily derived:
\begin{equation}\label{eq:density}
	h(x) = H(x)\left\{p\prime(x)\log F(x) + p(x)\frac{f(x)}{F(x)} - p\prime(x)\log G(x) + (1-p(x))\frac{g(x)}{G(x)}\right\},\nonumber
\end{equation}
where $p\prime(x) = \displaystyle\frac{1}{(b-a)}f_{\text{Beta}}\left(\frac{x}{(b-a)} \mid c_1,c_2\right)$ and $f_{\text{Beta}}$ is the density of the Beta distribution with shape parameters $c_1$ and $c_2$. Also,
\begin{equation*}
    f(x) = \frac{1}{\xi}F(x)z_1(x)^{-(1+1/\xi)}z_1\prime(x),\qquad g(x) = G(x)t_2(x)z_2\prime(x),
\end{equation*}
with $z_1\prime(x) = (\ell_{1-\beta/2,\xi} - \ell_{\beta/2,\xi})/s_\beta$ and $z_2\prime(x) = (\ell_{\beta/2} - \ell_{1-\beta/2})/\tilde{s}_\beta.$
Note that $f$ and $z_1\prime$ are defined for $x>q_\alpha - s_\beta(\ell_{1-\beta/2,\xi} - \ell_{\beta/2,\xi})^{-1}\ell_{\alpha,\xi}$.

\subsection{First and second derivatives of the bGEV density}\label{ap:derivbgevdensity}
Let 
$$m(x) = p\prime(x)\log F(x) + p(x)\frac{f(x)}{F(x)} - p\prime(x)\log G(x) + (1-p(x))\frac{g(x)}{G(x)}.$$
The we can write $h(x) = H(x)m(x) $ and
\begin{align*}
    h\prime(x) &= h(x)m(x) + H(x)m\prime(x),\\
    h\prime\prime(x) &= h\prime(x) m(x) + 2h(x)m\prime(x) + H m\prime\prime(x),
\end{align*}
with
\begin{align} 
 	m\prime &= p\prime\prime\{\log F - \log G\} 
	+ 2p\prime\left\{\frac{f}{F} - \frac{g}{G}\right\}
	+ p\left(\frac{f\prime F - f^2}{F^2}\right)
	+ (1-p)\left(\frac{g\prime G - g^2}{G^2}\right),
	&\nonumber\\
	&\nonumber\\
	m\prime\prime &= p\prime\prime\prime\{\log F - \log G\} 
	+ 3p\prime\prime\left\{\frac{f}{F} - \frac{g}{G}\right\}
	+ 3p\prime\left\{\frac{f\prime F-f^2}{F^2} - \frac{g\prime G-g^2}{G^2}\right\}+\nonumber\\
	&p\left\{\frac{f\prime\prime F^2 - 3ff\prime F + 2f^3}{F^3}\right\}
	+ (1-p)\left\{\frac{g\prime\prime G^2 - 3gg\prime G + 2g^3}{G^3}\right\}.\nonumber
\end{align}
Note that we removed the $x$ argument for ease of notation. Note also that

\begin{equation*}
    p\prime\prime(x) = \frac{1}{(b-a)^2}f\prime_{\text{Beta}}\left(\frac{x-a}{b-a}\mid c_1,c_2\right),\quad
    p\prime\prime\prime(x) = \frac{1}{(b-a)^3}f\prime\prime_{\text{Beta}}\left(\frac{x-a}{b-a}\mid c_1,c_2\right),
\end{equation*}
where
{
\begin{align*}
    f\prime_{\text{Beta}}(y\mid c_1,c_2) &= \frac{(c_1-1)y^{c_1-2}(1-y)^{c_2-1} - (c_2-1)y^{c_1-1}(1-y)^{c_2-2}}{\beta(c_1,c_2)},\\
    f\prime\prime_{\text{Beta}}(y\mid c_1,c_2) &= \frac{(c_1-1)\{(c_1-2)y^{c_1-3}(1-y)^{c_2-1}-(c_2-1)y^{c_1-2}(1-y)^{c_2-2}\}}{\beta(c_1,c_2)}\\
    &- \frac{(c_2-1)\{(c_1-1)y^{c_1-2}(1-y)^{c_2-2}-(c_2-2)y^{c_1-1}(1-y)^{c_2-3}\}}{\beta(c_1,c_2)}.
\end{align*}
}
Finally, the first two derivatives of $f$ and $g$ can be expressed as
{\footnotesize
\begin{align*}
    f\prime(x) &= \frac{z_1(x)^{-(2+1/\xi)}z_1\prime(x)}{\xi}\{f(x)z_1(x) - (1+1/\xi)F(x)\},\\
    f\prime\prime(x) &= \frac{z_1(x)^{-(3+1/\xi)}z_1\prime(x)}{\xi}[-(2+1/\xi)\{f(x)z_1(x)-(1+1/\xi)F(x)\} + z_1(x)\{f\prime(x)z_1(x) + f(x)(z_1\prime(x) - 1 -1/\xi)\}],\\
    g\prime(x) &= g(x)t_2(x)z_2\prime(x) - G(x)t_2(x)(z_2\prime(x))^2,\\
    g\prime\prime(x) &= g\prime(x)t_2(x)z_2\prime(x) - 2g(x)t_2(x)(z_2\prime(x))^2 - G(x)t_2(x)(z_2\prime(x))^3.
\end{align*}}

\baselineskip 16pt
\bibliographystyle{CUP}
\bibliography{references.bib}

\end{document}